\documentclass[prb,twocolumn,amsmath,showpacs]{revtex4-1}
\usepackage{color}
\usepackage{graphicx}
\usepackage{mathrsfs}
\usepackage{ulem}

\usepackage{ifpdf}
\ifpdf
   \usepackage[unicode=true,colorlinks=true]{hyperref}
\fi

\renewcommand{\vec}[1]{{\bf #1}}

\newcommand{\bra}[1]{\left\langle {#1} \right\vert}
\newcommand{\ket}[1]{\left\vert #1 \right\rangle}
\newcommand{\brakt}[2]{\left\langle #1 \middle\vert #2 \right\rangle}

\begin{document}

\title{Fine structure of neutral acceptor states of isolated impurity in zinc-blende semiconductors}

\author{M.O.~Nestoklon}
\affiliation{Ioffe Physical-Technical Institute, Russian Academy of Sciences, St. Petersburg 194021, Russia}
\affiliation{CNRS-Laboratoire de Nanophotonique et Nanostructures, 91460 Marcoussis, France}
\author{O.~Krebs}
\author{R.~Benchamekh}
\affiliation{CNRS-Laboratoire de Nanophotonique et Nanostructures, 91460 Marcoussis, France}
\author{P.~Voisin}
\affiliation{CNRS-Laboratoire de Nanophotonique et Nanostructures, 91460 Marcoussis, France}

\begin{abstract}
  The properties of neutral acceptor states in zinc-blende semiconductors are 
  re-examined in the frame of extended-basis $sp^3d^5s^*$ tight-binding model. 
  The symmetry discrepancy between envelope function theory and atomistic 
  calculations is explained in terms of over symmetric potential in current 
  {\bf k$\cdot$p} approaches. 
  Spherical harmonics decomposition of microscopic Local Density Of States (LDOS) allows
  for the direct analysis of the tight-binding results in terms of envelope function.
  Lifting of degeneracy by strain and electric field and their effect on LDOS is examined. 
  The fine structure of magnetic impurity caused by exchange interaction of hole with 
  impurity $d$-shell and its dependence on strain is studied. 
  It is shown that exchange interaction by mixing heavy and light hole makes the
  ground state more isotropic.
  The results are important in the context of Scanning Tunneling Microscopy (STM) images 
  of subsurface impurities.
\end{abstract}


\maketitle

\section{Introduction}
Impurity states in semiconductors have been a major topic for the last 50 years,
due to combined fundamental interest and technological importance. In
particular, from the theoretical point of view, the related breaking of crystal
translational invariance is a tough problem. A prominent milestone is the
celebrated article of Luttinger and Kohn\cite{Luttinger55} where general {\bf k.p}, 
or Envelope Function
Theory (EFT) for electrons and holes in ``perturbed crystals'' and conditions
for decoupling interband and intraband matrix elements of the perturbating
potential were discussed. An oversimplified view has emerged that impurity
potential can be decomposed into a ``gentle'' potential having no interband
matrix elements, and a ``central cell correction'' associated to details of the
potential near the impurity center. In the seventies, the case of neutral
acceptors and the role of valence band degeneracy was thoroughly discussed by
Lipari and Baldereschi.\cite{Baldereschi73,Baldereschi74}
In short, even in absence of a central cell correction,
acceptor states are characterized by a spinor wavefunction with contributions
from the heavy, light and possibly split-off valence bands, and just as the host
valence band at the zone center, they have a four-fold degeneracy (or 
three fold
degeneracy if spin-orbit interaction is neglected). Generally speaking, EFT
approach aims at discarding the local, rapidly varying part of the wavefunction
and at establishing an effective hamiltonian acting on supposedly slowly varying
envelopes. Perhaps for this reason, in spite of its rigorous formalism, it tends
to favor approximations that discard the hardly tractable bond-length scale.
For instance, the obvious fact that, due to discrete positions of first
neighbors, charge distribution near the impurity center cannot be spherical was
essentially ignored, except for a lonely work of Castner\cite{Castner08,Castner09}
where the tetrahedral central cell correction $V_{tet}(\vec{r})$ is introduced 
to explain the properties of donors in Si. To our knowledge, the role of 
$V_{tet}(\vec{r})$ for acceptors has not been examined in the literature.

In parallel to these seminal theoretical contributions, a huge number of
spectroscopic and transport experiments were dedicated to studies of impurity
states in all kind of semiconductors, so that the topic was considered as
finished by the end of the seventies. However, there was recently a strong
revival of the interest in impurity physics due to the observation of single
impurities by scanning tunneling microscopy. The completely unexpected, strongly
non-spherical shapes of images associated with acceptors have given rise to
passionate debates about their supposed relation to impurity local density of
states.\cite{Jancu08,Yakunin04}
One of the most intriguing situation was the neutral acceptor state
associated with substitutional Mn in GaAs, because in addition to its deep
acceptor state character, it carries magnetic properties associated with
antiferromagnetic coupling between the ``weakly'' bound hole and the 5 electrons
occupying the 3$d$ shell of the Mn atom. It was assumed that the ``butterfly''
shape of Mn STM images was predominantly due to this magnetic character, until
very similar images were obtained in the case of GaP:Cd.\cite{Kort01} Cd in GaP gives a
neutral acceptor state with a binding energy in the 100 meV range, close to Mn
in GaAs, but obviously has no magnetic interactions. In these two examples, a
strong « central cell correction » is involved, since the binding energy is
about 4 times larger than that of a purely coulombic state. The corresponding
impurity radius is of the order of a nm, and it was soon realized that atomistic
models like tight binding were better suited than {\bf k.p} theory for modeling such
impurity states. In particular, many papers have relied on the simple $sp^3$ 
model with nearest neighbor interactions because it is supposed to give a 
fair account of valence band properties. In this paper, we
use the extended basis $sp^3d^5s^*$ model, which is known 
for giving accurate description of single particle states in semiconductors.

The paper is organized as follow: in section \ref{sec:symmetry}, 
we explain the fundamental symmetry mistake of current {\bf k.p} approaches;
in section \ref{sec:calcul} we compare different tight-binding 
models, study the difference between shallow and deep acceptors and 
introduce a spherical harmonic decomposition which gives us a powerful 
tool for an accurate qualitative and quantitative analysis of results of 
tight-binding calculations.
In section \ref{sec:strain}, we examine the lifting of acceptor 4-fold
degeneracy by strain fields.  
In section \ref{sec:E} we examine the lifting of 
degeneracy by external electric field.
In section \ref{sec:exchange} we study the role of exchange interaction and 
compare two approaches which are used in description of exchange, valid in 
two limit cases: for single non-interacting impurities and for a 
semimagnetic alloy.
In section \ref{sec:STM} we briefly discuss the 
scanning tunneling microscopy images of acceptors.
Section \ref{sec:conclusion} concludes the main results.
Appendices \ref{sec:Ylm} and \ref{sec:TB_spin} give some details of calculations.

\section{Symmetry mistake in current envelope function theories}\label{sec:symmetry}

For decades, the main tool for qualitative description of semiconductor micro-
and nanostructures was the {\bf k.p} theory. However, there are many examples
that without special care the {\bf k.p} approximation fails to reproduce some
basic properties even qualitatively. A large set of such failures are connected
with oversymmetrizing the problem considered. Examples are heavy-light hole
mixing at the interfaces,\cite{Ivchenko96, Krebs96}
$\Gamma-X$ mixing,\cite{Ando89,Ivchenko93}, 
valley \cite{Boykin04_APL,Boykin04_VO,Jancu04,Poddubny12}
and spin\cite{Nestoklon06, Nestoklon08} splitting induced by the interfaces,
etc.

The same problem arises when one wants to describe properties of a substitutional 
impurity in the {\bf k.p} framework.\cite{Schmidt07}
Naive approach using Luttinger 
Hamiltonian\cite{Luttinger55} together with Coulomb potential and a 
short-range potential representing\cite{Bernholc77,Lipari80} the central-cell
correction (accounting for chemistry of the impurity) fails to reproduce some
qualitative properties, due to the incorrect symmetry of the problem.\cite{Castner08,Castner09}
Indeed, Luttinger
Hamiltonian has the $O_h$ point symmetry\cite{Baldereschi73} and the Coulomb potential
has full rotational $O(3)$ symmetry. However, for a substitutional impurity this potential
is not centered at the inversion center of $O_h$ in diamond lattice,
but at an atomic site 
of the zinc-blende (or diamond) lattice, that only has the $T_d$ point symmetry. 
Therefore, independently of Cation/Anion inversion asymmetry, the substitutional 
impurity problem has intrinsic $T_d$ symmetry (which excludes inversion center),
and accurate treatment of the impurity must take into account this reduced 
symmetry. For a purely Coulomb potential, symmetry reduction would appear due 
to tetrahedral (octupolar) charge distribution on nearest 
neighbors,\cite{Castner08,Castner09} while for an iso-electronic center it 
would be linked to symmetry of impurity chemical bonding.
However, both in $O_h$ symmetry and $T_d$ symmetry the ground level of the
hole has $\Gamma_8$ symmetry and four-fold degeneracy. Symmetry reduction
changes the spacial structure of the wavefunctions, but it is not reflected
by additional energy splitting of the levels. This explains why the
symmetry mistake in current {\bf k.p} theories (that was actually mentioned
in Ref.~[\onlinecite{Baldereschi73}]) has not been revealed long ago
in experimental investigations.
Finally, it is worth noting that the deeper the impurity level is, the larger
quantitative effect of symmetry reduction will be. Obviously, the merit of
atomistic approaches (like atomistic pseudo-potentials or tight-binding) in
this context is that they automatically include correct symmetries.

\section{Tight-binding calculations}\label{sec:calcul}

Here we focus on tight-binding (TB) calculations of neutral acceptor
states (or valence-type iso-electronic centers) and first explain the
importance of using an extended-basis tight binding model.
It is often believed that the simple $sp^3$ TB model gives a fair account
of semiconductor valence band, while more complex schemes like the $sp^3d^5s^*$
TB model become necessary only when details of the conduction band zone
edge valleys come  into  play. However, it has been proved that the restricted-basis
$sp^3$ nearest neighbor TB model cannot account
quantitatively for the valence-band dispersion of III-V semiconductors,\cite{boykin99}
in contrast to the $sp^3d^5s^*$ model.\cite{boykin04, Jancu98}
Precisely, for GaAs the Chadi  $sp^3$ parameters give Luttinger parameters
$\gamma_1=5.37$, $\gamma_2=0.90$, $\gamma_3=1.81$, which are rather far from the
$sp^3d^5s^*$ values $\gamma_1=7.51$, $\gamma_2=2.18$, $\gamma_3=3.16$, the
latter being in excellent agreement with experimental results. These differences mean
erroneous effective masses and an underestimate of valence band warping in the
$sp^3$ model. In fact, there are large differences for valence band dispersion throughout the Brillouin
zone. It follows that the kinetic energy part of the impurity hamiltonian (about
100 meV for the deep neutral acceptor GaAs:Mn) is not correctly calculated in
the $sp^3$ model. The difference may be estimated separately for effective Bohr radius 
$a_0=\hbar^2\varepsilon_0\gamma_1/e^2m_0$ (33\% difference),
strength of the spherical spin-orbit interaction estimated from dimensionless 
coefficient\cite{Baldereschi73} $\mu=(6\gamma_3+4\gamma_2)/5\gamma_1$
(31\% difference) and cubic contribution\cite{Baldereschi73} 
$\delta=(\gamma_3-\gamma_2)/\gamma_1$, which is 26\% different in two models.

Also, it is worth to mention that without the empty $d$-orbitals, it is 
impossible to account quantitatively for the strain dependence
of band structure even in bulk semiconductors.\cite{Jancu98, Jancu07, Niquet09,Zielinski12}

\subsection{Shallow and deep acceptors}\label{sec:acceptors}

The properties of the acceptor states in {\bf k.p} approach are usually obtained by
adding to the Coulomb potential the central cell correction which may be 
used in different forms,\cite{Bernholc77,Lipari80} but generally it is 
some short-range potential which accounts for chemical properties of the 
impurity atom. 
In the tight-binding approximation central cell correction naturally comes as 
(i) a valence band offset of the virtual material which contains impurity atom 
(for instance Mn cation) and a counterpart (e.g. As anion)
and (ii) change of band structure of the host material due to strain field 
near impurity.
Strictly speaking, there is no freedom in choice of the tight-binding parameters 
as parameters of this virtual ``impurity'' material (which is actually a metal)
should be also fitted to 
{\it ab initio} and/or experimental data. However, for the sake of 
simplicity and taking into consideration that the states are not particularly 
sensitive to most of the tight-binding parameters of a single atom, we 
model ``general'' acceptor by modifying tight-binding parameters of the matrix 
adding only artificial valence band offset. 
It is worth to note that accurate tight-binding treatment with modification of 
tight-binding parameters\cite{Jancu08}
implies change of parameters not only of the impurity
atom, but also of the nearest neighbors, and two-center hopping integrals connecting them.
Since the main effect of this procedure is the renormalization of 
diagonal energies, in the following we adopt a simplified approach 
of adding a potential similar to that used in Ref.~\onlinecite{Bhattacharjee00}:
\begin{equation}
U(r) = \frac{e^2}{\varepsilon r} + U_{cc} \exp \left(-\frac{r^2}{a_{cc}^2} \right).
\end{equation}
This approach allows us to study how hole state localized at the 
acceptor behaves by changing very few parameters of central cell correction 
and neglecting details of the chemical structure of impurity. 
In the following we will use a fixed central cell correction radius 
$a_{cc}=2.5$\AA.

\begin{figure}[h!]
  \includegraphics[width=\linewidth]{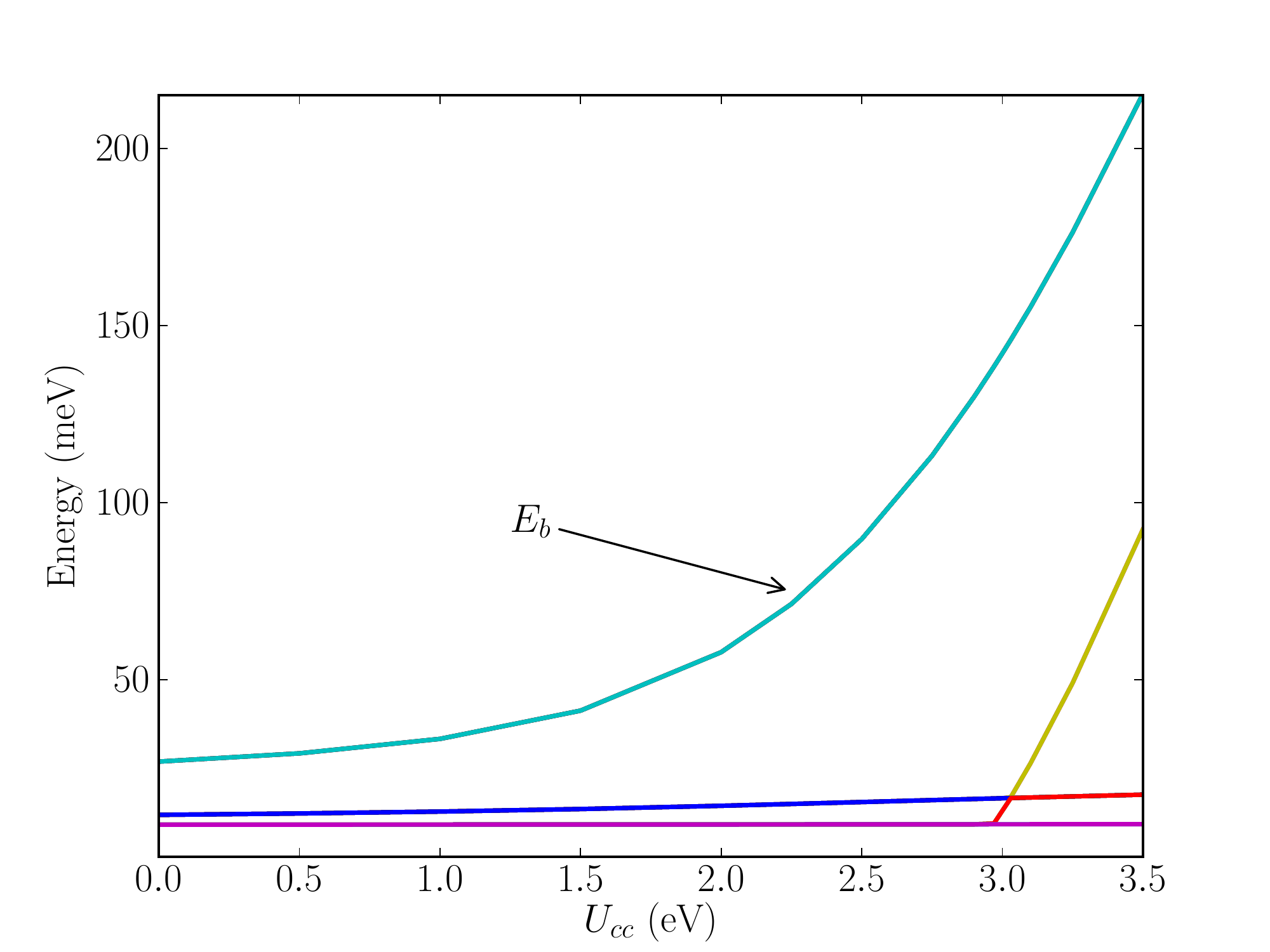}
  \caption{ 
    Binding energy for a neutral acceptor state in GaAs as a function of central cell 
    correction. One may note strongly non-linear behavior. For small 
    central cell correction, binding energy due to Coulomb 
    potential remains almost unchanged, but starting from some value
    it starts to increase rapidly. Above $U_{cc}= 3$~eV a new level associated with the split-off band shows up.
  }\label{fig:en_ccc}
\end{figure}

We consider GaAs as a prototype semiconductor, and unless opposite specification, 
acceptor states associated with substitution of a group II element 
on a cation site. The binding energy ($E_b$) as a function of 
central cell correction is presented in Fig.~\ref{fig:en_ccc}. 
One may see that the dependence of $E_b$ on central cell correction 
potential is highly non-linear. 
The change of binding energy due to central cell correction is proportional both to 
central cell correction $U_{cc}$ and amplitude of the wavefunction at the 
impurity. Increasing $U_{cc}$ we increase both, so the dependence of the 
binding energy on $U_{cc}$ is non-linear.

The calculation here is made using a rather large (262 144
atoms which form a cube with edge $\sim18$~nm) supercell, 
in order to minimize the width of the ``impurity miniband''
formed due to the periodization of the problem. For the purely coulombic
case, we have in the $\Gamma -X$ direction a residual bandwidth 
less than $0.9$~meV,
to be compared with the $26.88$~meV average binding energy. This large supercell
allows the calculation of acceptor excited states 
(which crudely may be labelled ``2S'' and ``2P'')
also displayed in Fig.~\ref{fig:en_ccc}. The widths of these excited state
minibands remain quite small (approximately $2.15$~meV for first two excited levels),
because they are
confined by topological minigaps, and their calculated binding energies are
reasonable. Yet, it is clear that the excited eigenstates are not as reliable
as the ground state.
Finally, the quantitative results of Fig.~\ref{fig:en_ccc}
depend on the whole set of material tight-binding parameters, in particular
those defining the relative weights of anion and cation contributions in
valence band Bloch functions.
For more ionic materials, the weight of anions is stronger and the 
relative importance of central cell correction on first neighbors 
in comparison with substitutional cation site is increased.

\begin{figure}[h!]
  \includegraphics[width=\linewidth]{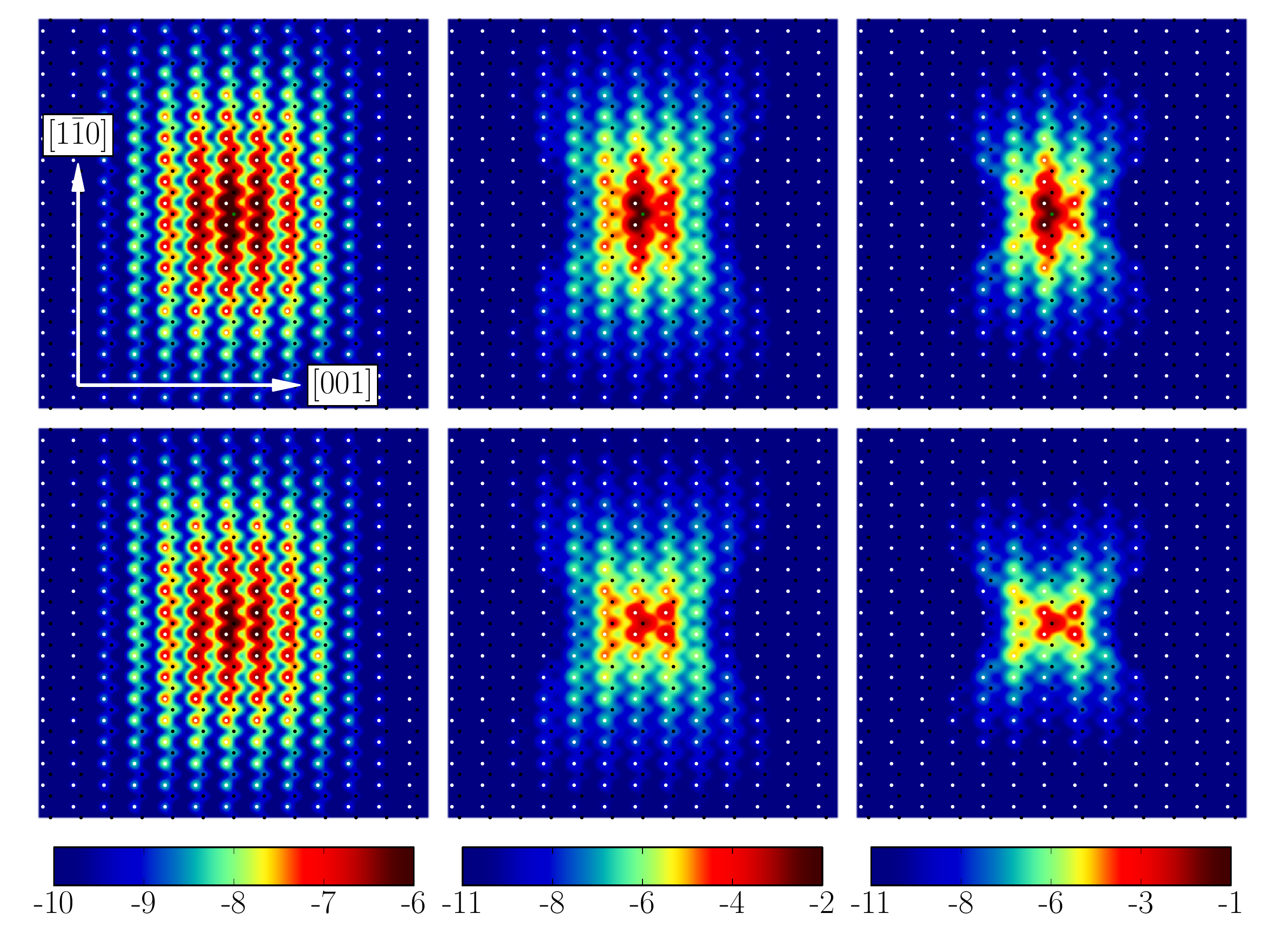}
  \caption{ 
    (110) plane cross sections of the impurity state LDOS 
    (log scale) for some characteristic cases: from left to right, 
    Coulombic center with binding energy $E_b=26.88$~meV,
    acceptor with $E_b=89.70$~meV,
    and $E_b=215.4$~meV.
    Note different color scales. States still have fourfold degeneracy. 
    Atoms are indicated with black (Ga) and white (As) dots.
    Upper raw for the wavefunction in impurity plane, lower for
    4-th atomic plane above.
  }\label{fig:en_ccc_wf}
\end{figure}
The wavefunction of the acceptor hole state for the case of 
shallow ($U_{cc}=0$~eV, $E_{b} = 26.88$~meV), 
intermediate ($U_{cc}=2.5$~eV, $E_{b} = 89.7$~meV), and 
deep ($U_{cc}=3.5$~eV, $E_{b} = 215.4$~meV)
acceptors are shown in 
Fig.~\ref{fig:en_ccc_wf} to illustrate how sensitive 
the shape of the wavefunction to the binding energy. 
Even for shallow acceptor, there is pronounced difference 
between $001$ and $110$ directions which stems from the valence 
band warping.
It is well seen that upon increasing binding energy, initially close 
to spherical state starts to feel the microscopic structure of the 
zinc blende lattice and becomes more and more ``tetrahedral''. 

\subsection{Spherical harmonic decomposition}\label{sec:flm}

While extremely helpful from a qualitative point of view, the type of
visualization used in Fig.~\ref{fig:en_ccc_wf} does not
provide a quantitative tool for wavefunction analysis, which is of utmost
importance when one gets interested in trend effects of perturbations
or change in parameters.

Note that through the rest of the manuscript, we will sometimes use term 
``wavefunction shape'' instead of Local Density Of State (LDOS). To avoid 
confusion, it is worth to mention that actual spinor wavefunctions are never 
plotted because most of the figures correspond to degenerate levels for which 
the choice of wavefunctions inside Hamiltonian eigensubspace is not unique.

To allow for the desired quantitative analysis, we performed the fit of tight-binding 
wavefunctions obtained in calculations with {\bf k.p}-like wavefunction as a 3D 
decomposition in spherical harmonics.

To perform successful fit, we first smooth tight-binding amplitudes by dressing  
them with gaussians. 
\begin{equation}\label{eq:n_TB}
  n^{TB}(\vec{r}) = \sum_{\alpha} |C_{i\alpha}|^2 
  \frac{e^{-(\vec{r}-\vec{r}_i)^2/a^2}}{(\sqrt{\pi}a)^3}
\end{equation}
where $C_{i\alpha}$ are the tight-binding coefficients at 
$i$-th atom and $a\simeq1.73$~\AA{} is chosen to be of the order of interatomic distance and 
$\vec{r}_i$ is the position of $i$-th atom.

Then we fit the smoothed amplitude \eqref{eq:n_TB} with a function 
\begin{equation}\label{eq:Ylm_fit}
  n(\vec{r}) = \sum_{l,m} f_{lm}(r) Y_{lm} (\Theta,\phi)
\end{equation}
using the freely available software archive SHTOOLS (shtools.ipgp.fr).
Since we are interested in the description of LDOS amplitudes, that are real, 
here we use real harmonics rather than complex spherical harmonics. 
The real harmonics are defined in Appendix~\ref{sec:Ylm}.

\begin{figure}[h!]
  \includegraphics[width=\linewidth]{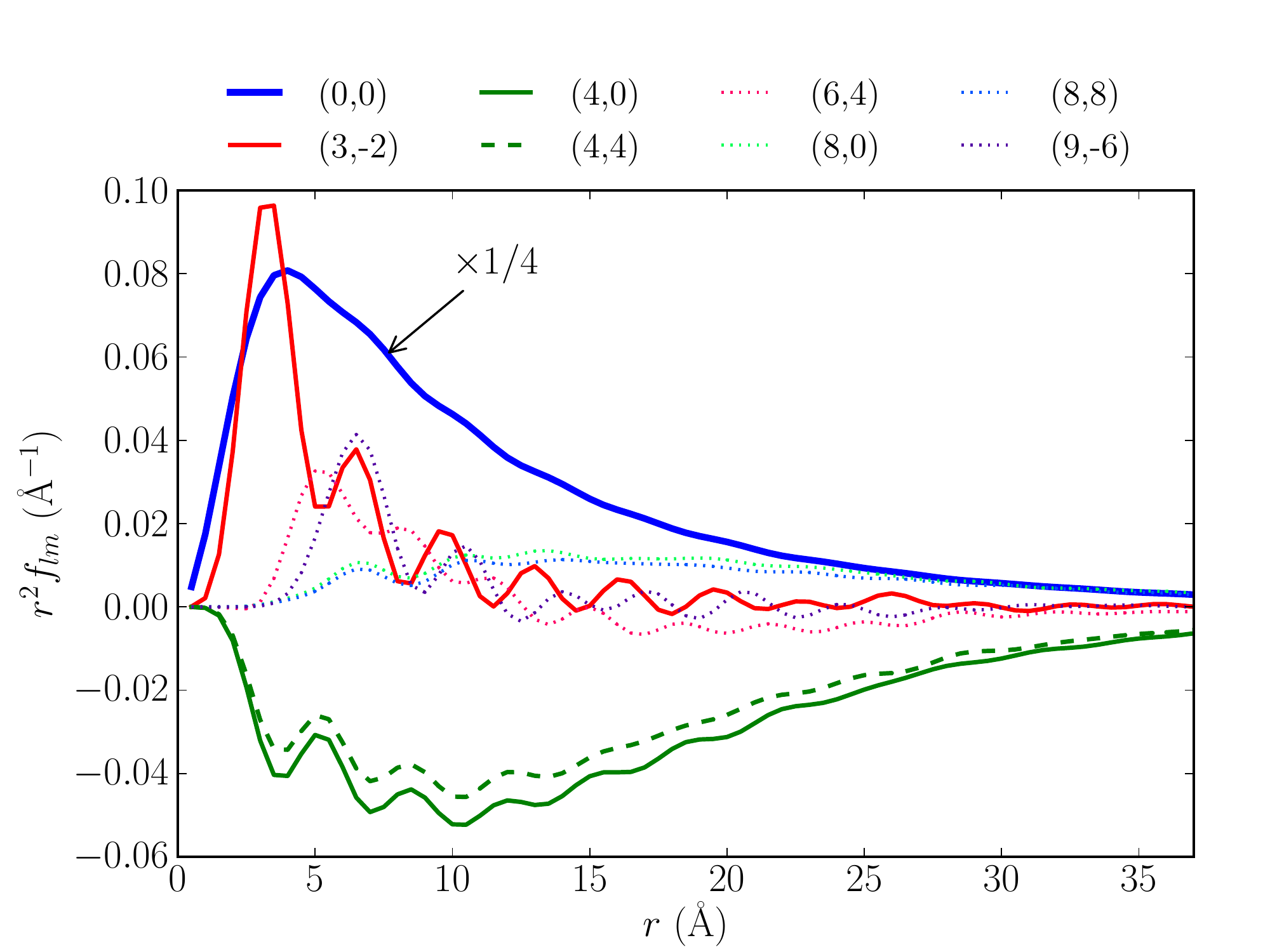}
  \caption{ 
    Functions $r^2 f_{lm}(r)$ for some harmonics. One may see that dominating 
    contributions come from 0-th and 4-th harmonics.
    The dominant contribution $r^2 f_{00}(r)$ is scaled with a factor $1/4$.
    Significant amount of 3-rd harmonics may be attributed to tetrahedral
    arrangement of the first neighbours of the impurity.
  }\label{fig:flm}
\end{figure}

The set of functions $f_{lm}(r)$ allows a quantitative description of the shape 
of the tight-binding LDOS of the impurity ground state. More precisely, noting
that $\int r^2 f_{00}(r)$ is the probability to find electron in 
the region limited by the radius of integration, we propose to use the
functions $r^2 f_{lm}(r)$ to estimate how 
``important'' harmonics $Y_{lm}(\Theta,\varphi)$ is in the impurity wavefunction. 
Also, the $r^2$ factor helps revealing aspherical features that characterize the 
``shape'' of impurity LDOS at some distance from impurity center.
An example may be found in Fig.~\ref{fig:flm}, where spherical harmonics 
decomposition of the level with binding energy $E_b=89.7$~meV 
from Fig.~\ref{fig:en_ccc_wf} is shown.

As one might expect, dominant contribution for 1S-like hole level is for 
$l,m=0,0$ harmonics. Luttinger Hamiltonian through the warping 
term\cite{Baldereschi74} proportional to $\gamma_3-\gamma_2$ 
 introduces harmonics $l=4, m=0,\pm4$. Tetrahedral site 
potential\cite{Castner08,Castner09} adds the harmonic $l=3, m=-2$.
Above mentioned contributions together may combine and introduce higher 
harmonics. Later we will discuss the effect of strain which reduces the
symmetry and introduces additional terms in the harmonic 
decomposition \eqref{eq:Ylm_fit}. In such case, the main additional 
contribution has the same symmetry as the applied perturbation, see below.

Note the oscillatory behaviour of these functions in 
Fig.~\ref{fig:flm}: they correspond to atomic texture of the crystal. 
For a few first coordinate spheres the atoms are distributed at some fixed 
distances with gaps between them which produces these oscillations. Gaussians in 
\eqref{eq:n_TB} tend to smooth them out, but there is a compromise between 
smoothing atomic texture and showing variations of harmonics in space.
To further simplify the ``shape description'', the set of 
$r^2 f_{lm}(r)$ functions can be replaced by a set of numbers
$P_{lm}$ representing the integrals over some range of $r$, that we 
shall use for a rough estimation of LDOS anisotropy:
\begin{equation}
  P_{lm} = \int_{r_1}^{r_2} r^2 f_{lm}(r)
\end{equation}
where in the following we choose $r_1=6$ \AA{} and $r_2=30$ \AA{}
to concentrate on a region without few 
neighboring atoms where validity of spherical harmonics 
decomposition is questionable and one better use atomistic 
results directly. For example, in the case of Fig.~\ref{fig:flm}, 
we have $P_{00}=2.42$, $P_{40}=-0.804$, $P_{44}=-0.69$ and $P_{80}=0.238$.

\begin{figure}[h!]
  \includegraphics[width=\linewidth]{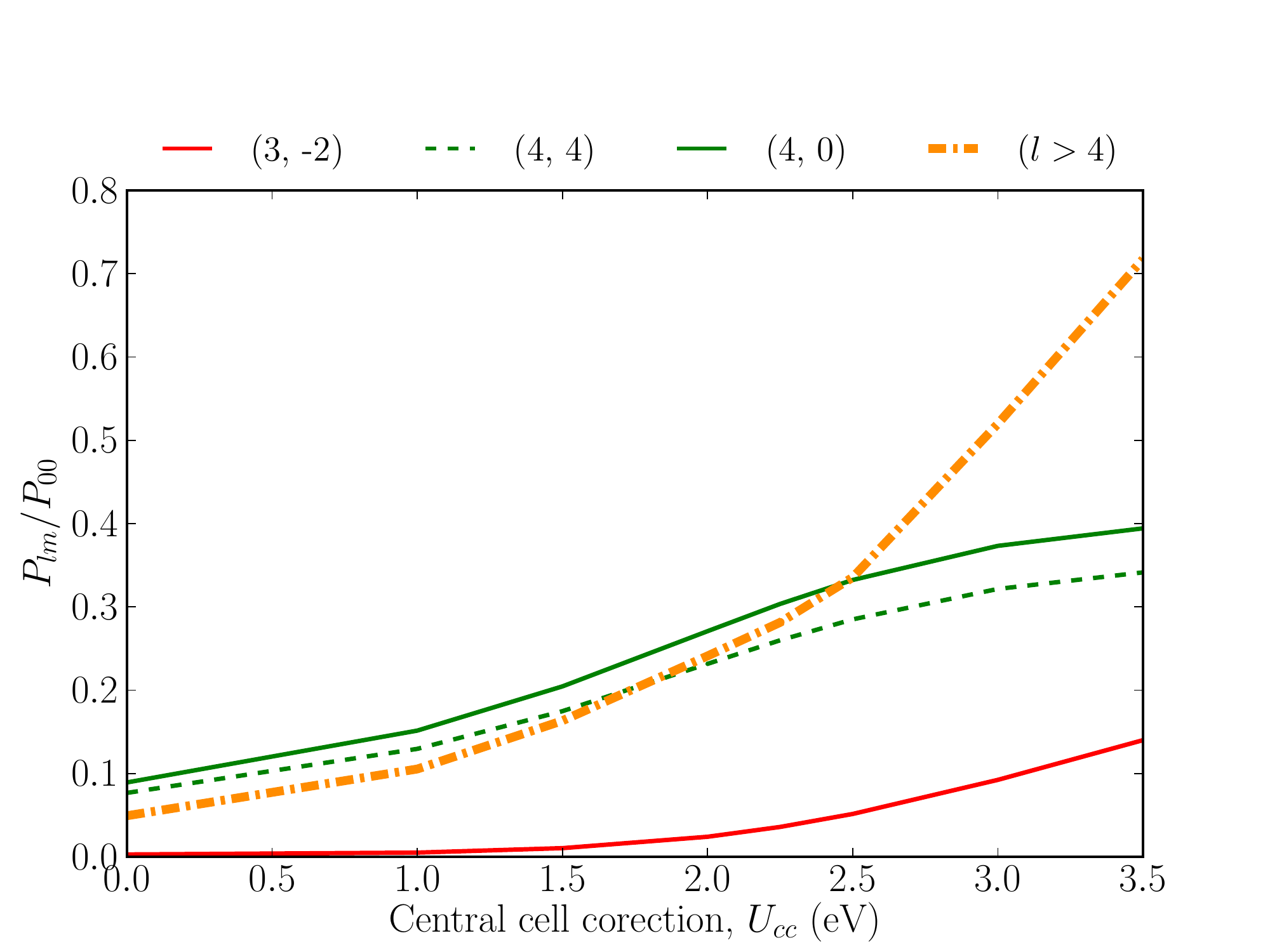}
  \caption{ 
    Absolute value of normalized coefficients $|P_{lm}/P_{00}|$ as a function of central cell 
    correction. Harmonics with $l>4$ are summed.
    It can be seen that 4-th harmonics which originates from terms with cubic symmetry
    in Luttinger Hamiltonian dominates and rises with the binding energy,
    until wavefunction becomes ``too localized'' and 
    higher harmonics start to dominate.
  }\label{fig:Plm_ccc}
\end{figure}

It is worth to note that cubic anisotropy has a maximum 
for binding energy around $100$~meV. For a Coulomb center, the kinetic 
energy is small and spherically symmetric contribution dominates. 
Increasing central cell correction and binding energy, 
we increase importance of cubic contribution $P_{40}$ and $P_{44}$, up to the moment when  
wavefunction becomes strongly localized and higher harmonics prevail, see Fig.~\ref{fig:Plm_ccc}.

\section{Effect of strain, analysis of impurity state}\label{sec:strain}
As long as the impurity potential respects $T_d$ symmetry, 
the hole ground state of the hole retains $\Gamma_8$ symmetry and remains four-fold 
degenerate, just as in the classical, oversymmetric {\bf k.p} approach. 
Time-inversion symmetry for a half-integer spin means that in absence of magnetic field 
and exchange interaction levels remain two-times degenerate. 
As a result, a perturbation respecting time-inversion symmetry
(like strain, quantum confinement, electric field...) can only lead to splitting
of one fourfold degenerate level into two Kramers-degenerate doublets.  In this
section, we use the spherical harmonics decomposition method and examine how
different uniaxial strains change the ground state LDOS.

The effect of strain is included into tight-binding following a generalized
approach of Ref.~\onlinecite{Jancu07} similar to one given in Ref.~\onlinecite{Niquet09}: the transfer 
matrix elements are scaled based on bond-length change and diagonal energy shift 
and splitting are introduced, based on a local strain tensor. This approach is shown to give a quantitatively 
correct description of strain in $sp^3d^5s^*$ tight-binding 
approximation.\cite{Jancu07,Zielinski12}

\subsection{Strain along [001]}\label{sec:strain_001}
When strain along [001] is applied, symmetry of the lattice is reduced from 
$T_d$ to $D_{2d}$ and four times degenerate level with symmetry $\Gamma_8$ 
is split in two levels $\Gamma_6\oplus\Gamma_7$ in accordance with compatibility table of the representations of this group.\cite{Koster}
Spherical harmonics decomposition shows how strongly the shape of hole 
density changes when the strain is applied. 
\begin{figure}[h!]
  \includegraphics[width=\linewidth]{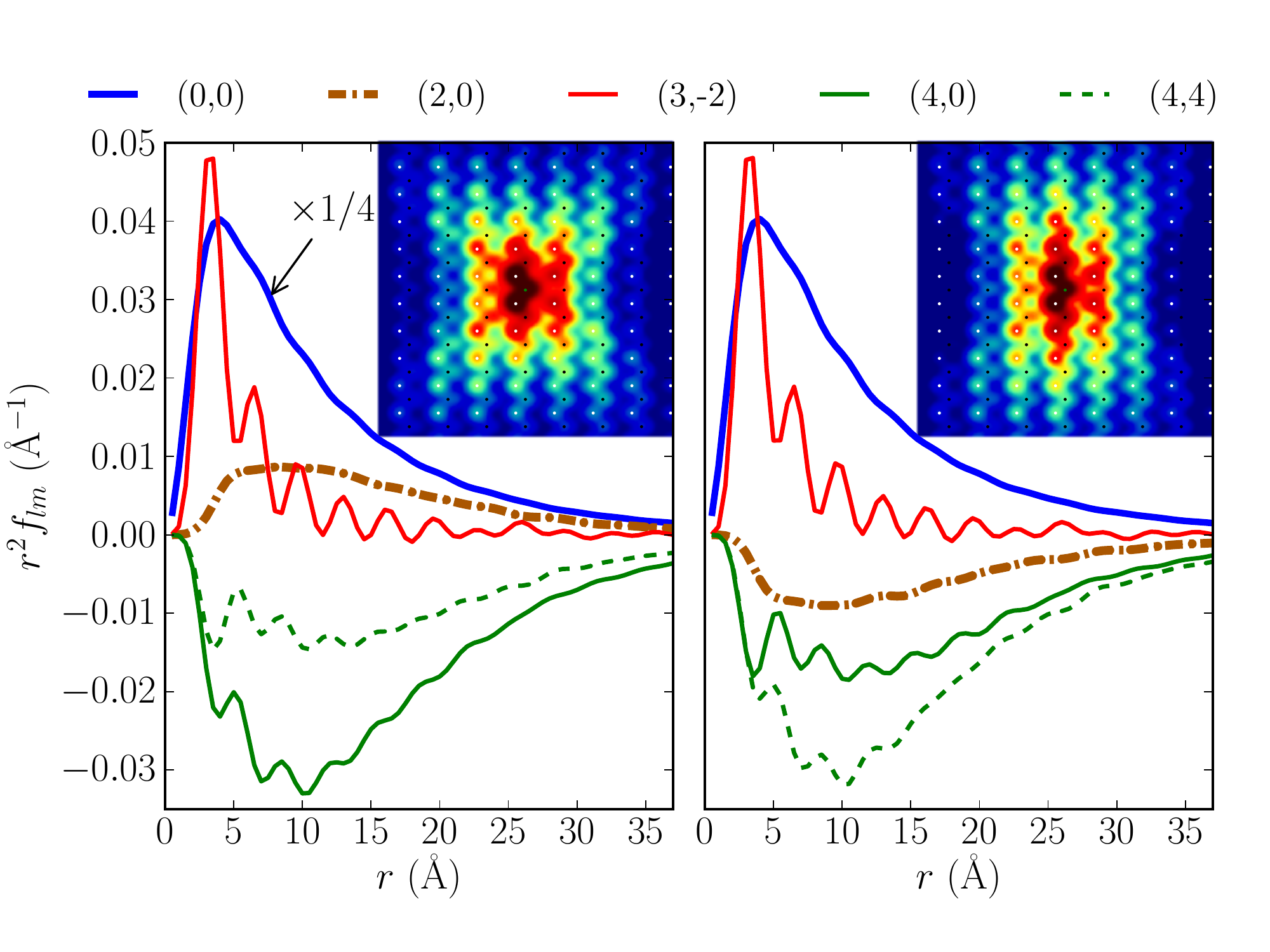}
  \caption{ 
    Comparison of dominant functions $r^2f_{lm}(r)$ for the two levels obtained after 
    ground state splitting by a weak [001] strain. 
    Strain value is $0.1$\%, and central cell correction is $2.5$~eV. 
    Left panel shows level with binding energy $88.58$~meV and right panel 
    shows level with binding energy $91.08$~meV.
    For the left figure $P_{20}/P_{00}=0.109$ and for the right one 
    $P_{20}/P_{00}=-0.115$.
    Note that summing of the amplitudes for the two doublets, one gets
    almost the same amplitudes as for the unstrained, fourfold-degenerate case.
    Binding energies are arbitrarily counted from the top of unstrained 
    GaAs valence band.
  }\label{fig:flm_001}
\end{figure}

The change in the $P_{lm}$ values is quite pronounced. It is pretty obvious 
that the four-times degenerate ground state which looks more or less like cube 
is split by the strain to two levels: one is elongated along strain and the other 
is flattened in that direction. These changes are reflected in the additional 
$P_{20}$ and in the related changes in  $P_{40}$ and $P_{44}$

The figures show that the spherical harmonics decomposition gives an adequate 
numerical criterion to study this change of the shape.

\subsection{Strain along [110] and [111]}\label{sec:strain_110}

\begin{figure}[h!]
  \includegraphics[width=\linewidth]{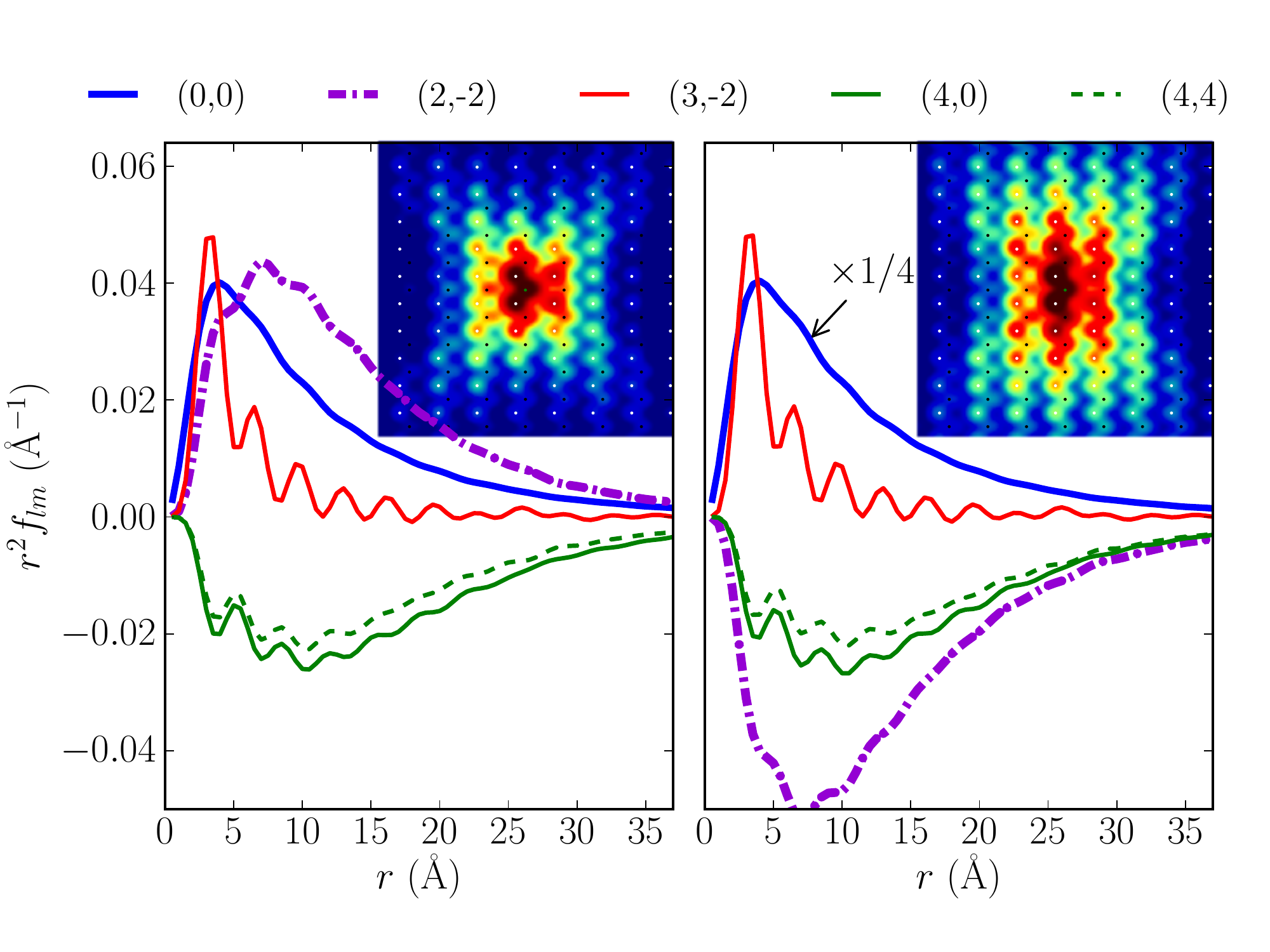}
  \caption{ 
    The same as in Fig.~\ref{fig:flm_001}, but strain along [110].
    Left panel shows level with binding energy $84.77$~meV 
    and right panel shows level with binding energy $95.52$~meV.
    For the left figure $P_{2-2}/P_{00}=0.431$ 
    and for the right one $P_{2-2}/P_{00}=-0.526$.
  }\label{fig:flm_110}
\end{figure}

Even though the symmetry is lower in the case of both  [110] and [111] strain directions
in comparison with [001] strain, without magnetic field we cannot split fourfold 
degenerate level more than in two Kramers doublets and the energy dependence on the 
strain applied is rather poor. 
The spherical harmonics decomposition however allows carefully examine how exactly 
the shape of the two levels changes. From Fig.~\ref{fig:flm_110}
one may see that again second harmonics of the significant value and opposite sign
appears for the two levels. 

Similar results may be obtained for the strain along [111]. 

Obviously, the splitting of neutral acceptors states by uniaxial strain reflects 
the corresponding splitting of valence band edges. For instance, the larger 
splitting for [110] than for  [001] strain is linked with the larger 
value of the deformation potential $d \sim -5$~eV coming into play in the 
former case compared to $b \sim -2$~eV governing the latter. However, it is 
important to note that bound state splitting is significantly smaller 
than band-edge splitting\cite{Bir63,Schairer74,Morgan75}
(twice smaller in given example), because "heavy" and "light" holes 
remain admixed in the split states.

\section{Effect of external electric field}\label{sec:E}
Another interesting case of symmetry-breaking perturbation is the effect of an
electrostatic potential. External electric field is known to cause a quadratic
Stark shift of hydrogenic impurity ground state, but at the same time it
reduces the point group symmetry of the acceptor problem. For instance, electric
field along a [001] axis reduces the symmetry from $T_d$ to $C_{2v}$, which implies
an admixture of $(3/2, 3/2)$ and  $(3/2, -1/2)$ components in the spinorial
wavefunction. Calculation shows that in addition to the quadratic Stark shift a
linear splitting occurs. For a binding energy of $100$~meV, the wave-function
polarizability is weak, and both the linear splitting and quadratic shift have
a similar value ($0.3$ and $0.25$~meV respectively)
for $F=100$~kV/cm. From the symmetry considerations, we expect
qualitative differences when changing the electric field orientation with respect
to cubic symmetry axis. Details will be published elsewhere.

Decomposition in spherical harmonics shows expected appearance of dipole
momentum which is associated with coefficient $(l=1,m=0)$ and reflects the 
symmetry of the perturbation. This dipole component $P_{10}/P_{00}$
of a sum over four levels 
is linear with electric field (about 5\% for $100$~kV/cm) while the dipole 
momentum for the split levels is enhanced (decreased) to about 3\%.

\section{The case of magnetic acceptor}\label{sec:exchange}
So far we considered the effect of perturbations that break the $T_d$ symmetry
and lift the fourfold degeneracy of the neutral acceptor state. In this
section, we consider the more subtle situation of the perturbation by exchange
interaction with $d$-electrons, like in  the emblematic case of GaAs:Mn.
By itself, Heisenberg coupling between bound hole and the $d$-shell electrons
does not reduce the symmetry, but it increases the size of the Hilbert space
and mixes angular momenta of the components into total angular momentum of 
the system.
The related many-body interaction completely changes the impurity spectrum by
introducing a spin dependent fine structure. Here, we compare the fine structures
obtained using either the isotropic (Heisenberg) or axial (Ising) exchange
couplings, and calculate how they are influenced by a $(001)$ strain. A striking
result is that isotropic exchange  makes ground state LDOS more resistant against
the strain-induced symmetry breaking.

In the following we will extensively use the notation $3/2$ and $5/2$ 
while for the exchange interaction we adopt spherically symmetric
approximation. Two microscopic sources of reduced symmetry may be 
considered: crystal field splitting of the states of half-filled Mn 
$d$-shell and anisotropic exchange. Both are allowed by $T_d$ symmetry 
and may split excited states.
However, existing experimental measurements\cite{Bihler08} show that these
splittings are extremely small 
compared with spherically symmetric part and we neglect them completely.

A number of available experimental
data\cite{Schneider87,Astakhov08,Kudelski07,Krebs09,Krebs13}
shows that the ground state of hole 
localized at Mn acceptor is three times degenerate.
Adopting isotropic model, it is interpreted as hole state with angular
momentum $3/2$ interacting with Mn $d$-shell with angular momentum $5/2$ which
gives four states with total angular momenta $\mathscr{F}=1,2,3,4$.

\begin{figure}[h!]
  \includegraphics[width=0.9\linewidth]{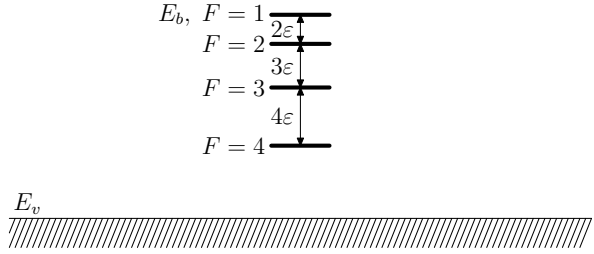}
  \caption{
    Scheme of the Mn impurity levels. In the isotropic approximation 
    the ground level of single impurity has total angular momentum $F=1$.
  }\label{f_levels}
\end{figure}

It is well established that exchange interaction in single 
$\{$Mn + hole$\}$ complex
is relatively small. Some authors estimate the value $\varepsilon$ which is a 
half splitting between ground and first excited level to be of the order of 
$5$~meV,\cite{Linnarsson97,Averkiev87,Averkiev88} some report even smaller 
values about $1$-$2$~meV.\cite{Sapega00,Sapega01,Sapega02}
There are also papers\cite{Bhattacharjee00} about the determination of 
the parameter $N_0\beta$ used to describe $p$-$d$ exchange in GaMnAs materials, 
but the link to the 2-spin exchange ($\epsilon$) depends on the wavefunction
of the acceptor state on the $A^-$ core. In the model of  Bhattacharjee (see below)
the typical value for $N_0\beta$ would agree with epsilon about 5 meV.
In the following we will consider a value 5meV. 
This allows us, following Bhattacharjee\cite{Bhattacharjee00}, to start from calculations of the 
ground state of hole bound to acceptor neglecting exchange interaction 
(but of course with account on spin-orbit splitting) and then add exchange 
interaction. 
This scheme is valid due to the fact that the binding energy $E_b\simeq100$~meV is 
significantly larger than exchange interaction splitting $\varepsilon\simeq 5$~meV. 
Schematic view of the levels structure is given in Fig.~\ref{f_levels}. 

Wavefunction of this composite system may be obtained using standard 
procedure:\cite{Varshalovich}
\begin{equation}\label{eq_explicit_momentum}
  \ket{F,m} = \sum_{|m_1|\le \frac32} C^{Fm}_{\frac52m-m_1,\frac32m_1} 
              \ket{\frac52,m-m_1}\ket{\frac32,m_1},
\end{equation}
  where $C^{Fm}_{Jm_2,Sm_1}$ are Clebsh-Gordan coefficients.
  We will not use \eqref{eq_explicit_momentum} explicitly because numerically it is 
  easier to obtain the same result as a diagonalization of exchange Hamiltonian operator 
  matrix. 

  The Hamiltonian of the problem reads as 
  \begin{equation}\label{eq_Hamiltonian_def}
    \mathcal{H}=\mathcal{H}_0+\mathcal{H}_{ex}
  \end{equation}
where $\mathcal{H}_0$ is the kinetic energy plus impurity potential and 
$\mathcal{H}_{ex}$ is the exchange interaction. We first solve the problem
with $\mathcal{H}_0$ exactly in the tight-binding approximation and 
afterwards diagonalize full Hamiltonian of the system
$\mathcal{H}$ in the truncated basis of four function 
which form the ground state. This follows the general scheme proposed in 
Ref.~\onlinecite{Bhattacharjee00}, the discussion of validity of this 
approximation will be given later.

Without strain, the solution of $\mathcal{H}_0$ is the
four times degenerate hole level which is in terms 
of $\vec{k}\cdot \vec{p}$ approximation $1S_{3/2}$ hole level.\footnote{
We again remind that in tight-binding approximation this level has $\Gamma_8$ 
symmetry and not $D_{3/2}$}
\begin{equation}\label{eq_ground_state_0}
  \mathcal{H}_0 \ket{1S_{3/2},i} = E_{b} \ket{1S_{3/2},i}\;\;i=1,2,3,4
\end{equation}
Including strain, this four times degenerate level splits in two. 
Due to time-inversion symmetry, the remaining two-times 
degeneracy holds for arbitrary strain.
Applying magnetic field one may in principle fully split the ground state in 
four levels. In the following we do not rely on exact ground state degeneracy.
\begin{equation}\label{eq_ground_state}
  \mathcal{H}_0 \ket{1S_{3/2},i} = E_i \ket{1S_{3/2},i}
\end{equation}

Following Bhattacharjee\cite{Bhattacharjee00} we add the exchange 
interaction in the simplest form 
\begin{equation}\label{eq_exchange}
  \mathcal{H}_{ex}= - \mathscr{J}({\vec{r}}) \vec{J} \cdot \vec{S},
\end{equation}
where $\vec{S}$ is the spin operator acting in Mn spin configuration space and 
$\vec{J}$ is the spin operator acting on hole wavefunction. 

Then we consider exchange interaction in the truncated basis which is obtained as 
a Cartesian product of the ground state of the hole \eqref{eq_ground_state} 
and spin states of Mn. 
\begin{equation}\label{eq_exchange_coord}
  H_{li,kj}^{ex} =\bra{\frac52,l} \bra{1S_{3/2},i} \mathcal{H}_{ex} \ket{\frac52,k} \ket{1S_{3/2},j}.
\end{equation}
Hole wavefuncion in the tight-binding approach is a sum over atoms
\begin{equation}
  \ket{1S_{3/2},j} = \sum_{n\alpha} c^{j}_{n\alpha} \ket{\vec{r}_n,\alpha}.
\end{equation}
In the $sp^3d^5s^*$ model $\alpha$ runs through 20 basis orbitals.\cite{Jancu98}

As long as exchange interaction is local, it is natural to assume that 
$\mathscr{J}(\vec{r})$ is nonzero at Mn atom only.
Under this assumption \eqref{eq_exchange_coord} reduces to 
\begin{equation}\label{eq_exchange_tb}
  H_{li,kj}^{ex} = - \mathscr{J}(0)\sum_{\alpha\beta} c^{i*}_{0\alpha} c^{j}_{0\beta} 
     \bra{\frac52,l} \bra{r_0,\alpha} \vec{J}\cdot\vec{S} \ket{\frac52,k} \ket{r_0,\beta },
\end{equation}
where we start enumeration from Mn.
This equation reduces to 
\begin{equation}\label{eq:exch_tb}
  H_{li,kj}^{ex} = - \mathscr{J}(0)\sum_{\gamma=1..3} \{ S^{5/2}_{\gamma} \}_{lk} 
     \sum_{\alpha\beta} c^{i*}_{0\alpha} \{ J_{\gamma} \}_{\alpha\beta} c^{j}_{0\beta}.
\end{equation}

Here $S_{\gamma}^{5/2}$ are standard spin matrices of the total momentum $5/2$ and 
$J_{\gamma}$ are matrices of the spin operator in the tight-binding basis
\begin{equation}\label{eq_spin_def}
   \vec{J}_{\alpha\beta} =  \bra{r_0,\alpha} \vec{J} \ket{r_0,\beta }.
\end{equation}

To write explicit form of the matrix \eqref{eq_exchange_coord} we need to compute 
matrix elements of the spin operator. 
Detailed derivation of it is out of the scope of current paper, we
comment this procedure briefly in appendix~\ref{sec:TB_spin}. 

The wavefunctions with account on exchange are then found as a 
solution of eigenproblem 
\begin{equation}
    \sum_{kj}\left[ E_i\delta_{ij}\delta_{lk} + H^{ex}_{li,kj}  \right]\xi_{kj}^{\mathscr{F}} = 
    \epsilon_{\mathscr{F}} \xi_{li}^{\mathscr{F}},
\end{equation}
where $\mathscr{F}$ enumerates both total spin of the state $F$ and its projection $m$.
Energies of the levels are $\epsilon_{\mathscr{F}}$ and the wavefunctions are
\begin{equation}
    \ket{\mathscr{F}} = \sum_{ij} \xi_{ij}^{\mathscr{F}} \ket{1S_{3/2},i} \ket{\frac52,j}.
\end{equation}
Or, explicitly in the tight-binding basis,
\begin{equation}
    \ket{\mathscr{F}} = \sum_{n\alpha}  \left( \sum_{ij}\xi_{ij}^{\mathscr{F}} c^{i}_{n\alpha}\ket{\frac52,j} \right), \ket{\vec{r}_n,\alpha} 
\end{equation}
from which one may easily compute tight-binding amplitude of the hole wavefunction 
\begin{equation}\label{eq:density}
    n(\vec{r}) = \brakt{\mathscr{F}}{\mathscr{F}} = \sum_{n\alpha} 
    \left( \sum_{ijl} \xi_{ij}^{\mathscr{F}*}\xi_{lj}^{\mathscr{F}}
    c^{i*}_{n\alpha}c^{l}_{n\alpha} \right) \brakt{\vec{r}_n\alpha}{\vec{r}_n\alpha}.
\end{equation}

Note that the scheme used here is valid also in a small magnetic field or in extreme 
case of very dilute alloy. However, when the splitting due to external or effective 
magnetic field is larger than exchange interaction of single Mn+hole complex 
($\simeq 5$~meV), classical spin description\cite{MacDonald_09,TAMR_APL} 
should be used instead. 

\begin{figure}[h!]
  \includegraphics[width=\linewidth]{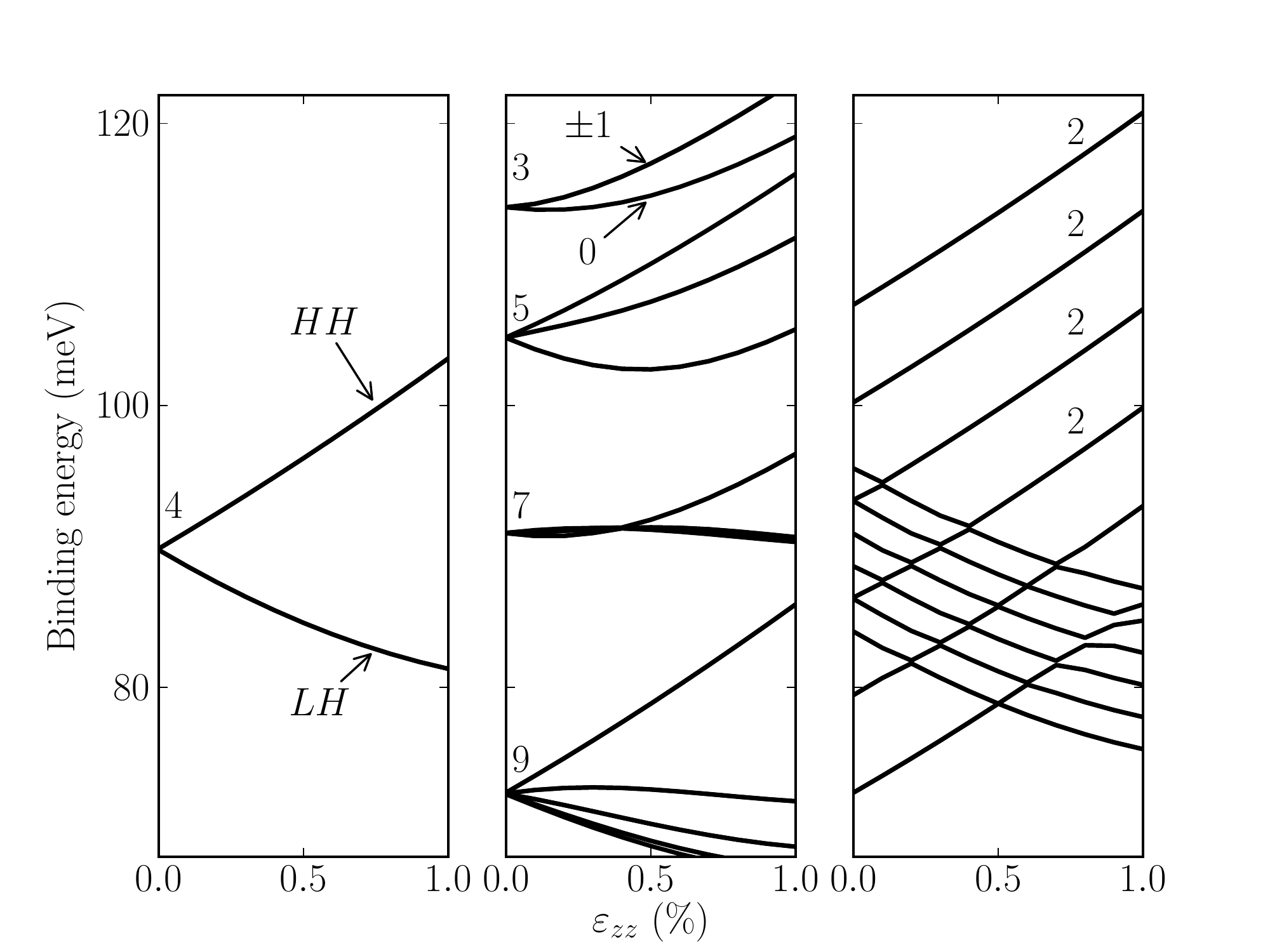}
  \caption{ 
    Splitting of the $\Gamma_8$ hole ground state without (left panel) and 
    with (center panel) account on exchange interaction with Mn $d$-shell.
    Strain along [001] axis.
    For comparison, we also show results if we coinsider exchange operator as 
    $\mathcal{H}_{ex}^{I} = J_z \otimes S_{z}$
    Energy zero corresponds to top of valence band in unstrained GaAs.
  }\label{fig:StrSplit}
\end{figure}

In Fig.~\ref{fig:StrSplit} we present a structure of levels the hole ground 
state localized at Mn impurity neglecting exchange interaction 
completely (left panel); by using the scheme 
(\ref{eq_Hamiltonian_def}-\ref{eq:density}) (central panel); and 
considering Ising exchange with fixed classical vector $\vec{S}=(0,0,S)$ (right panel)
as a function of strain applied along $[001]$.

It is clearly seen that in semi classical Ising description which acts as a 
local magnetic field, the splitting added to each level is proportional 
to the product of Mn spin projection and hole spin projection. The distance 
between levels is independent on strain along [001] because such strain
only adds additional splitting between $\pm\frac32$ and $\pm\frac12$ states of 
the hole. All levels are two times degenerate because in this approach 
$E_{+s+j}=E_{-s-j}$ and $E_{-s+j}=E_{+s-j}$ (here $s$ and $j$ are respectively
Mn and hole spin projections).

In contrast, scheme (\ref{eq_Hamiltonian_def}-\ref{eq:density}) gives 
four levels which are three, five, seven and nine times degenerate in 
accordance with quantum mechanical angular momentum summation. Which is more 
important, this splitting is lifted by applying the strain to the structure 
because the states with different angular momentum projection have different
heavy/light hole ratio and feel their splitting by strain.

\begin{figure}[h!]
  \includegraphics[width=\linewidth]{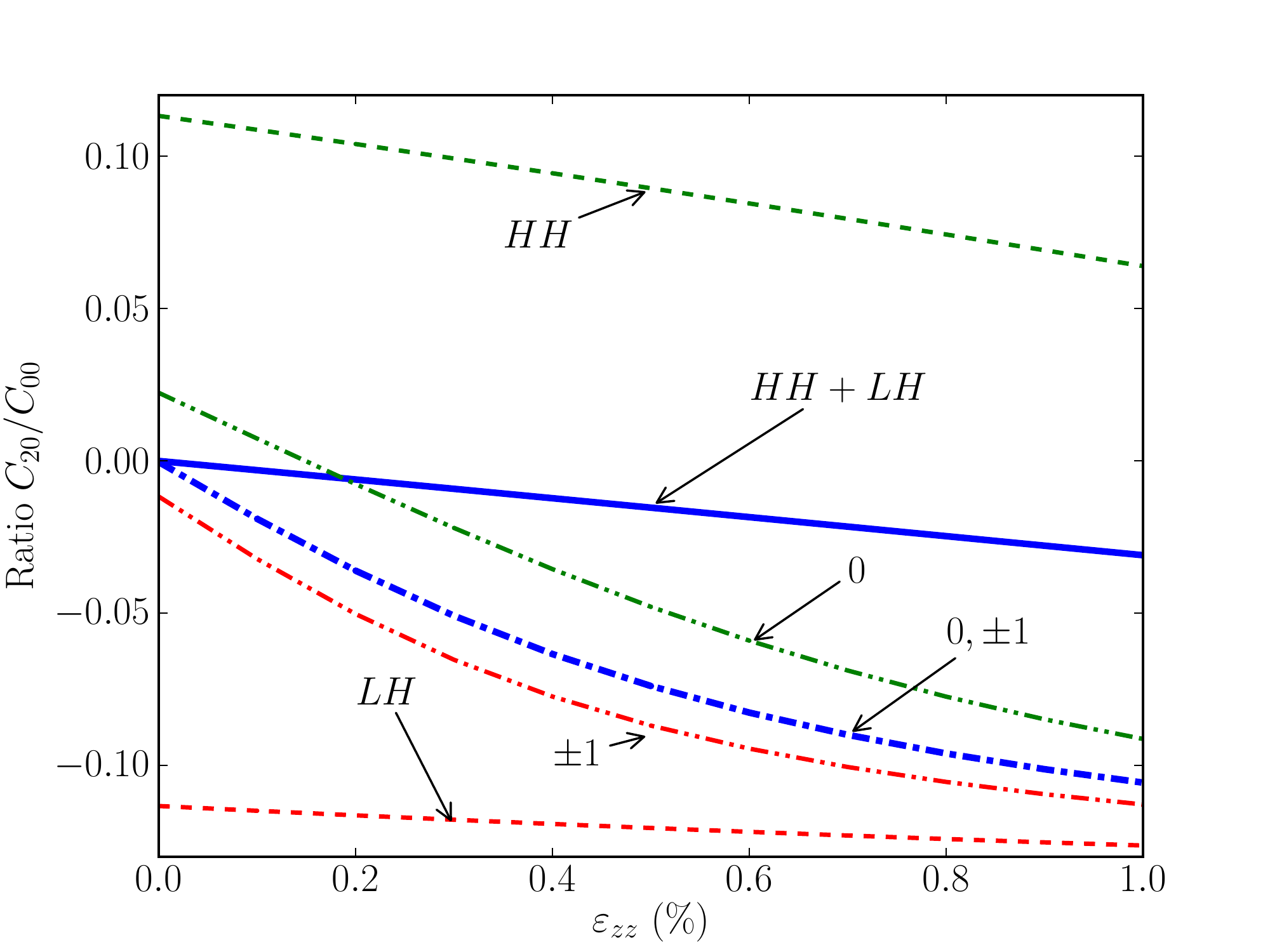}
  \caption{
  Ratio $P_{20}/P_{00}$ as a function of strain applied along [001].
  Solid line shows this ratio for a sum over four states which originate
  from $\Gamma_8$ ground state. Dash lines show this ratio for two dublets 
  in which ground state is split under strain.
  Dashed-dot and dash-dot-dot lines show anisotropy for the lowest in energy 
  levels with account on exchange interaction with impurity with angular 
  momentum 5/2.
  Dash-dot line shows it for the sum of three levels with total angular 
  momentum 1 and two dash-dot-dot lines show it for the levels which 
  may be associated with total angular momentum projection $0$ and $\pm1$.
  }\label{fig:C20}
\end{figure}

To analyze the interplay between exchange interaction and strain in more 
details, in Fig.~\ref{fig:C20} we show the ratio $P_{20}/P_{00}$ as a 
function of strain applied along $[001]$ direction calculated with and 
without account on exchange interaction. 
This ratio quantitatively shows the oblateness of the corresponding states.

From Fig.~\ref{fig:C20} it is clear that without exchange interaction 
spherically symmetric four-times degenerate level is split by the strain in 
two levels with opposite oblateness which may be attributed to heavy and light hole.
Exchange interaction by mixing heavy and light hole into three states with 
total angular momentum $1$ makes the ground state significantly more isotropic.
Which is important, the resulting levels shape is almost linear with the 
strain.

We would like also to comment the validity of the approximation 
(\ref{eq_Hamiltonian_def}-\ref{eq:density}) for the description of 
the exchange interaction. Eigenproblem with the Hamiltonian 
\eqref{eq_Hamiltonian_def}
may be solved exactly in the tight-binding framework, but this 
assumes the solution in product space which is rather expensive from the 
computational point of view. 
Effective potential for the ground state which comes from 
exchange interaction will be proportional to $\mathscr{J}(\vec{r})$ which
adds a contribution to the central cell correction. 
This correction depends on a spin configuration of the composite 
system and is seen as different central cell correction for different levels.
As long as the exchange interaction is small compared with localization energy, 
this correction is small compared with the main part and approach presented 
in this section may be 
used. Available experimental data show that it is the case of Mn acceptor. 
However, the effect of the exchange interaction for excited levels or for 
speculative magnetic impurity with similar exchange interaction and small 
binding energy demands for the exact solution.

\section{STM images}\label{sec:STM}
Observation of sub-surface neutral acceptor states by scanning tunneling
microscopy\cite{Yakunin04,Mahieu05,Loth06,Jancu08}
has been the origin of a considerable renewal of interest in
acceptor physics.\cite{Bihler08,Astakhov08,MacDonald_09,Flatte_PRB,flatte,yasdani}
In early papers,\cite{Yakunin04} the STM images
were compared with cross sections of bulk impurity LDOS in a (110) plane.
However, it appears that the situation is far more complex: indeed, 
in the vicinity of a $(110)$ surface there is a large elastic deformation 
known as the
``surface buckling''.\cite{Engels98}
Besides, the presence of the surface and the tip
induced bend-bending add a perturbation qualitatively similar to the effect
of an electric field acting on the hole. In addition, STM measures LDOS at some
distance in the vacuum and this implies hybridization of impurity and dangling-bond
states.\cite{Jancu08}
In Ref.~\onlinecite{Jancu08}, it was assumed that splitting by surface buckling
(about $40$~meV for an impurity in the fourth sublayer) was much larger than both
thermal energy and exchange splitting. The magnetic interaction was therefore
neglected, and it was assumed that only the ground state LDOS was observed in
low temperature STM imaging. These conclusions were also supported by the fact
that Zn and Cd acceptor in InP and GaAs,\cite{Kort01} which is non-magnetic, 
give images remarkably similar to Mn acceptor in GaAs. 
All thee impurities have very similar binding energy in the $100$~meV range.

Complete discussion of STM results in the
frame of present theory is beyond the scope of this paper, but we note that the
analysis presented above suggests that for a strain induced splitting of
$40$~meV, exchange would be negligible for the $\epsilon=5$~meV which is 
consistent with available experimental data.

\section{Conclusion}\label{sec:conclusion}
We have re-examined the theory of neutral acceptor states within the $spds^*$ 
extended-basis tight-binding model, that combines exact account of local symmetries
and high accuracy of band dispersion representation. A spherical harmonic decomposition
of the tight-binding LDOS has been used and allows both qualitative and quantitative
analysis of the numerical results. The lifting of acceptor fourfold degeneracy by
symmetry breaking perturbations like uniaxial strain or external electric field has
been explicated, as well as the acceptor fine-structure arising from exchange
interaction with $d$-electrons in the case of a magnetic impurity. This computational
approach can be further improved by considering a full TB parameterization of the
impurity central cell potential from comparisons with {\it ab initio} calculations of the
``impurity material'', as done in Ref.~\onlinecite{Jancu08} for GaAs:Mn. The same formalism
can provide realistic modeling of more complex situations like Mn-doped quantum dots,
impurity pairing, sub-surface impurities and their STM imaging, or acceptor states in
presence of external electric and magnetic fields.

\appendix

\section{Spherical harmonics}\label{sec:Ylm}

For completeness, we give a exact form of real spherical harmonics 
used in decomposition \eqref{eq:Ylm_fit}:
\begin{equation}
Y_{lm} (\Theta,\phi)
=
\begin{cases}
\bar{P}_{lm}(\cos\Theta)\cos m\phi & \mbox{if } m\ge 0\\
\bar{P}_{l|m|}(\cos\Theta)\sin |m|\phi & \mbox{if } m < 0
\end{cases},
\end{equation}
where the normalized associated Legendre functions are given by
\begin{equation}
\bar{P}_{lm}(t)=\sqrt{(2-\delta_{0m})(2l+1)\frac{(l-m)!}{(l+m)!}}P_{lm}(t)
\end{equation}
with the following definition of the associated Legendre functions:
\begin{align}
P_{lm}(t)&=(1-t^2)^{m/2}\frac{d^m}{dt^m} P_l(t),\\
P_l(t) &= \frac1{2^l l!} \frac{d^l}{dt^l} (t^2-1)^l.\nonumber
\end{align}

\section{Spin operator in tight-binding}\label{sec:TB_spin}
  We follow philosophy of group representation theory, and start from the definition 
  of state with angular momentum $s$ as a wavefunction which
  transforms under action of space rotations 
  under definite representation $\mathcal{D}_s^{\pm}$. 
  Technically, this definition cannot be transferred to tight-binding directly, because tight-binding theory is not 
  isotropic in accordance with space crystal symmetry. 
  However, the same arguments apply for $\vec{k}\cdot\vec{p}$ model where one still may work
  in the spherical approximation\cite{Baldereschi73} and associate angular momentum with the 
  states.

  To generalize this approach to tight-binding we note that any point group
  is a subgroup of the full rotation group and thus its representations may
  be chosen as a subgroups of its representations as well. 
  We choose basis function so that $\Gamma_6$, $\Gamma_7$ and $\Gamma_8$ are the subgroups of
  $\mathcal{D}_{1/2}^{+}$, $\mathcal{D}_{1/2}^{-}$, $\mathcal{D}_{3/2}^{-}$. 
  Accordingly, we associate angular momentum with representations of $T_d$. 
  This procedure should be used with care however, it only makes sense if one wants to 
  discuss the results of the tight-binding in terms of $\vec{k}\cdot\vec{p}$ model or to 
  transfer approximation used in $\vec{k}\cdot\vec{p}$ to TB and can 
  not be extended beyond envelope function approach without additional assumptions. 

  As a  result, we write spin operator as
  \begin{equation}\label{eq:TB_spin}
      \vec{J} = \sum_{i,\alpha} \alpha \ket{\Gamma_{i,\alpha}}\bra{\Gamma_{i,\alpha}},
  \end{equation}
  where $i$ enumerates representations of $T_d$ and $\alpha=\pm1/2$ for $i=\Gamma_6,\Gamma_7$
  and $\alpha=+3/2,+1/2,-1/2,-3/2$ for $i=\Gamma_8$. The functions $\ket{\Gamma_{i,\alpha}}$ 
  are constructed from tight-binding orbitals. E.g., $\ket{\Gamma_{6,+1/2}}=\ket{s}\uparrow$, etc.

  Approach \eqref{eq:TB_spin} might seem too formal at the first glance, however it provides two 
  essential properties one usually expects from the theory: 1) it is very formal,
  contains no explicit assumptions on basis wavefunctions or any other details of 
  the model and may be applied in any situation; 2) for the bulk states near band edges 
  in unstrained zincblende semiconductor it gives results which are usually expected from 
  the $\vec{k}\cdot\vec{p}$ theory, namely electron with spin $1/2$ and holes with total spin $3/2$ which 
  are split to heavy and light holes with spin projection to momentum direction $\pm3/2$ and 
  $\pm1/2$ respectively, etc.

\section*{Acknowledgments} This work was supported by ``Triangle de la Physique'' 
(CAAS project), by International Ioffe Institute-CNRS Associate Laboratory ILNACS and by
RFBR grants, and EU project SPANGL4Q
 
\bibliography{nac}

\end{document}